\documentclass[conference]{IEEEtran}
\IEEEoverridecommandlockouts
\usepackage{cite}
\usepackage{amsmath,amssymb,amsfonts}
\usepackage{algorithmic}
\usepackage{graphicx}
\usepackage{textcomp}
\usepackage{xcolor}
\usepackage{comment}
\usepackage{svg}
\usepackage{subcaption}

\def\BibTeX{{\rm B\kern-.05em{\sc i\kern-.025em b}\kern-.08em
    T\kern-.1667em\lower.7ex\hbox{E}\kern-.125emX}}
    
\usepackage{enumitem}

\setenumerate[1]{itemsep=0.5pt,partopsep=0pt,parsep=\parskip, topsep=2pt, leftmargin=12pt,}

\usepackage{etoolbox}
\makeatletter
\patchcmd{\@makecaption}
  {\scshape}
  {}
  {}
  {}
\makeatletter
\patchcmd{\@makecaption}
  {\\}
  {.\ }
  {}
  {}
\makeatother
\def\tablename{Table}    
    
\begin{document}

\title{A Cycle-Accurate Soft Error Vulnerability Analysis Framework for FPGA-based Designs

}

\author{\IEEEauthorblockN{Eduardo Rhod,~Behnam Ghavami,~Zhenman Fang,~Lesley Shannon}
\IEEEauthorblockA{School of Engineering Science\\
Simon Fraser University\\Burnaby, British Columbia, Canada\\
Emails: \{eduardo\_rhod,behnam\_ghavami,zhenman,lesley\_shannon\}@sfu.ca\\}}
\maketitle

\begin{abstract}
Many aerospace and automotive applications use FPGAs in their designs due to their low power and reconfigurability requirements. Meanwhile, such applications also pose a high standard on system reliability, which makes the early-stage reliability analysis for FPGA-based designs very critical. 

In this paper, we present a framework that enables fast and accurate early-stage analysis of soft error vulnerability for small FPGA-based designs. Our framework first extracts the post-synthesis netlist from an FPGA design. Then it inserts the bit-flip configuration faults into the design netlist using our proposed interface software. After that, it seamlessly feeds the golden copy and fault copies of the netlist into the open source simulator Verilator for cycle-accurate simulation. Finally, it generates a histogram of vulnerability scores of the original design to guide the reliability analysis.
Experimental results show that our framework runs up to 53x faster than the Xilinx Vivado fault simulation with cycle-level accuracy, when analyzing the injected bit-flip faults on the ITC'99 benchmarks. 

\end{abstract}


\vspace{-0.15in}
\section{Introduction}

An FPGA usually contains a set of configurable logic blocks (CLBs) and programmable interconnect resources, making it possible to implement flexible digital systems in many areas such as aerospace, automotive, machine learning, and datacenters~\cite{fpga-wiki}. In order to have high density and fast configuration speed, the majority of FPGAs use SRAM (Static RAM) technology to store the configuration bitstream. However, SRAM-based FPGAs are vulnerable to radiation induced soft errors \cite{8103796, 7336736}. Soft errors may occur with a particle strike to a sensitive region of a transistor in a SRAM cell resulting in an inversion of the logical value, which is called bit-flip. The bit-flips occurring in the FPGA configuration memory remain unchanged during the following cycles. Such bit-flips are permanent and can cause failure in an FPGA-based design. Therefore, it is very important to analyze such bit-flips in FPGA-based designs and provide corresponding protection and/or correction, especially for highly-reliable systems used in aerospace and automotive. 

\begin{figure}[!tb]
  \includegraphics[width=\columnwidth]{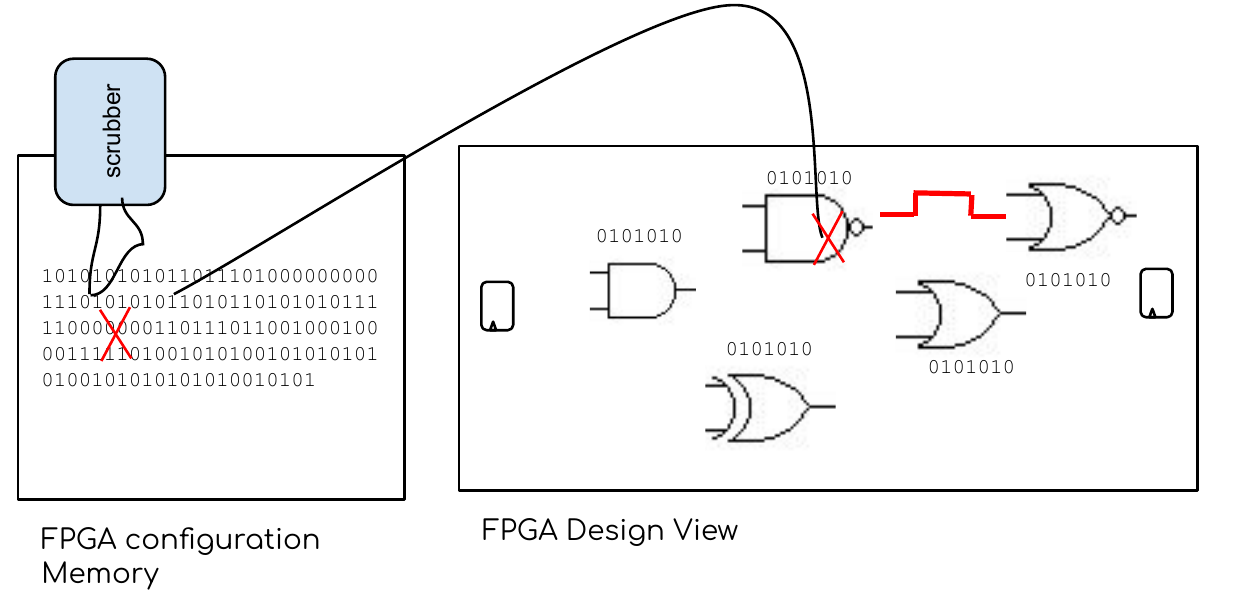}
  \caption{A configuration memory scrubber is a controller that detects a bit-flip and its location, and performs a partial reconfiguration for bit-flip correction.}
  \vspace{-0.1in}
  \label{fig:FPGA}
\end{figure}

Modern FPGAs often use error correction methods, such as scrubbing, to correct configuration memory bit-flips \cite{6412749,fpga-wiki}. As shown in Fig. \ref{fig:FPGA}, memory bit-flips are usually corrected using coding mechanisms, or golden models, together with partial reconfiguration \cite{6132703,7776929}. As a result, single event upsets (SEUs) need not be permanent and their consequences can be classified as a transient fault that lasts until a scrubbing occurs. Moreover, if a bit-flip effect is further masked in the circuit design, this error will not be propagated to the circuit outputs. And thus the effect of that bit-flip will be completely transparent to the system. Therefore, to implement a highly-reliable design that still achieves high performance, it is essential to precisely analyze which parts of the FPGA-based design are most vulnerable to soft errors (bit-flips) and thus provide lightweight error protection and/or correction.


As presented in Section~\ref{sec:related}, there are many methods developed to analyze the reliability of FPGA-based designs [9-48]. However, there is still the need for a fast and accurate toolflow, that is open source, to perform early-stage analysis of soft error vulnerability for FPGA-based applications in the design phase.


In this paper, we present a framework that enables fast and cycle-accurate soft error vulnerability analysis of an FPGA-based design using the post-synthesis netlist as input.
First, given an initial FPGA design or partial design in the early design stage, our framework extracts the post-synthesis netlist. Based on the extracted netlist, it automatically inserts the bit-flip configuration faults. Then it seamlessly feeds the golden copy and fault copies of the netlist into the open source simulator Verilator~\cite{OMS} to perform cycle-accurate simulation. According to the simulation results, it generates a histogram of vulnerability scores for each component in the original design.
To demonstrate the usefulness of our framework, we use it to analyze the reliability of the widely used ITC'99 benchmarks~\cite{ref:ITC99} with injected bit-flip faults and show which components are more vulnerable to bit-flip errors.
Compared to the Xilinx Vivado fault simulation, our framework runs up to 53x faster while achieving cycle-level accuracy. 



\begin{table*}[t]
\begin{center} \caption{Comparison of fault injection techniques for FPGA-based designs}
  \label{tab:1} \begin{tabular}[t]{c |c c c c c} \hline & Radiation Techniques & Run-time Reconfiguration & Fault Emulation & Analytical Modelling & \textbf{Fault Simulation} \\ [1ex]  \hline Controllability & low & fair & high & fair & \textbf{high}\\ \hline Run-time & ultra fast & fast & fast & fair & slow \\ \hline Accuracy & very high & high & high & fair& \textbf{high} \\ \hline Flexibility & low & fair & fair & fair & \textbf{high} \\
\hline Design flow stage & prototype & prototype & prototype & early& \textbf{early} \\ \hline Cost & high & fair & fair & low & \textbf{low} \\ \hline\end{tabular}
\vspace{-0.2in}
\end{center}
\end{table*}

\section{Review of Soft Error Analysis for FPGAs}
\label{sec:related}

To analyze the vulnerability of soft errors for an FPGA-based design, a common technique (called \emph{fault injection}) is to intentionally inject faults to disturb some parts of the design and analyze the possibility of changes of its normal behavior \cite{44380}. There are many different means to insert artificial faults into an FPGA-based design, and we classify them as follows.
\begin{enumerate}
    \item \emph{Using radiation techniques to induce upsets in actual physical FPGA design}~\cite{8103796, 10.5555/1302493.1302729, 8906534}: These methods are fast and realistic. But they need a prototype version of the design and cannot be applied in early stages of the design flow. The lack of controllability of fault locations is another weakness. In addition, these methods destroy the equipment and are quite expensive.
    
    \item \emph{Modifying configuration memory to force upsets}~\cite{1369494,5491825,4291792,8632000,8307104,OMS1}: Some studies try to modify an FPGA’s  configuration bitstream file directly, where others modify an FPGA’s configuration memory at run-time using dynamic partial reconfiguration.  According to the different configuration ports, these methods can be divided into two categories: external and internal fault injections. Note that they need the prototype version of the design.
    
    \item \emph{Emulating faults in FPGAs}~\cite{EBRAHIMI20141000,6555963,4515957,4677166,6131392}: These studies add some extra hardware (saboteurs) to the original design and run the modified version on the board to emulate the behaviour of bit-flips. Existing techniques in this subject manipulate the design’s description in RTL (Register-Transfer Level) or post-implementation netlist.
    These methods need complex internal or external components for controlling the fault injection and inference processes. Therefore, they either need dedicated external hardware platforms or impose considerable area overhead. 
    In addition, they  need the prototype version of the design.
    
    \item \emph{Analytical modelling that applies probabilistic and/or statistical analysis to model the behavior of fault masking in an FPGA design}~\cite{10.1145/2554688.2554767,6459610,1589186,6873692,1515755}: These methods often use worst case analysis. Hence they 
    usually lead to inaccurate results.
    
    \item \emph{Fault simulation that uses a software simulator to simulate the behaviour of memory upsets for a given design descried in hardware description languages (HDLs)}~\cite{robach2003simulation,gil2003study,4711556,1568808,6555963,6734986,7500787,614074,6154066,6850649,5325359,6879587}: These methods use either code modification to add mutants and saboteurs or simulator commands to change variable and signal values. Fault simulation offers important advantages over other techniques. They have both high observability and controllability of the faults, and can be applied soon at various level of design abstractions. Hence they are very useful for reliability aware CAD (computer-aided design) tool development. However, fault simulation methods suffer from long run-time.
\end{enumerate}
We summarize the fault injection techniques in Table \ref{tab:1} and compare their controllability, run-time, accuracy, flexibility, design flow stage, and cost. Considering the pros and cons of fault injection methods, fault simulation is a good choice to analyze the reliability of an FPGA design accurately at early design stages. On one hand, high-level fault simulation suffers from high inaccuracy compared to low-level fault simulation~\cite{7500787}. On the other hand, low-level fault simulation tools~\cite{6879587,6378210} use event-driven simulation, and thus, suffer from long execution time and large memory consumption.

In this paper, we explore a  cycle-accurate fault simulation framework to enable accurate yet fast soft-error vulnerability analysis for FPGA-based designs. We decide to explore the fault simulation at the post-synthesis netlist level, which can be considered precise because it provides accurate fault injection locations in an already optimized netlist. 

\section{Proposed Framework for Soft error Vulnerability Analysis}\label{sec:framework}

\begin{figure}
  \vspace{0.05in}
  \includegraphics[width=\columnwidth]{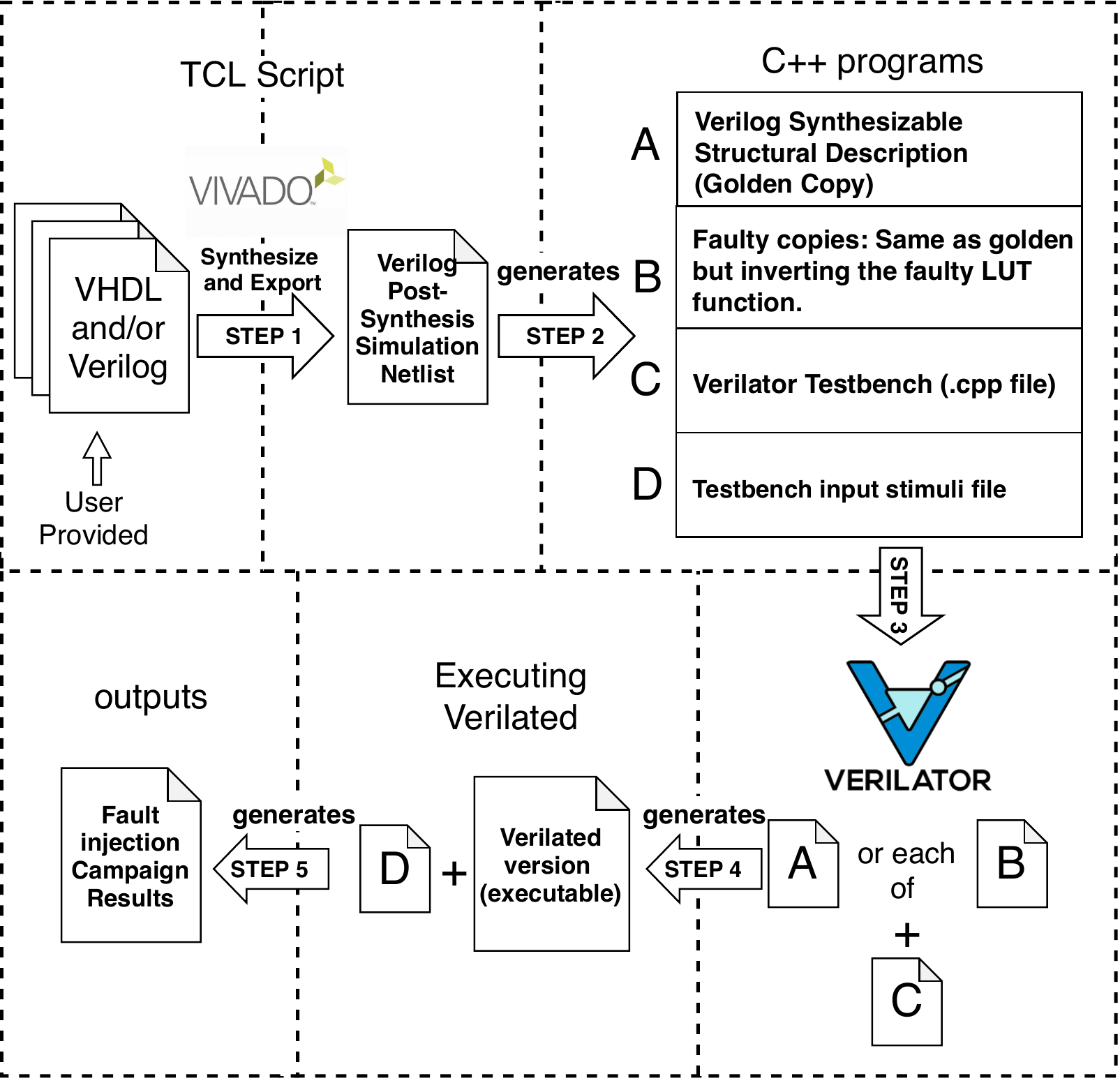}
  \caption{The proposed framework for soft error vulnerability analysis of an FPGA-based design.}
  \vspace{-0.1in}
    \label{fig:FLOW}
\end{figure}

The overall structure of our proposed framework is presented in Figure~\ref{fig:FLOW}. Given an input FPGA circuit described in VHDL or Verilog, it can automatically extract the post-synthesis netlist and insert bit-flips at the basic element (BEL) level, and seamlessly feed the golden and fault copies to the open source simulator Verilator \cite{OMS} for fast and cycle-accurate simulation. Finally, it will generate a histogram of vulnerability scores for the input circuit.
Next, we present more details of each component of our framework.

\subsection{Post-Synthesis Design Netlist Extraction}

During the synthesis stage, a digital circuit expressed in synthesizable VHDL or Verilog code is synthesized to a lower level Verilog netlist, with the TCL (Tool Command Language) command 'write\_verilog' in a TCL script to automate the process. This exported Verilog netlist describes a digital circuit in terms of basic elements (BELs) that can be used in a post-synthesis simulation of the circuit. It is a structural description that interconnects the circuit primitive modules (BELs), which are generated by the Xilinx Vivado synthesis tool for the targeted FPGA. Each BEL has its behavioural description defined in a Xilinx proprietary library intended for behavioural simulation purpose only. For each BEL, its proprietary library is then replaced by a synthesizable behavioural Verilog description, so that it can be fed to the open source simulator Verilator for cycle-accurate simulation. Note, that in this work, we use the flatten hierarchy property of the synthesis step and we have no DSP in the synthesis option as the behaviour of faults in DSP elements is out of scope in this paper. Our current post-synthesis netlist extraction is developed for the Xilinx toolflow only; but it can be easily extended for the Intel Quartus toolflow as well.

\subsection{Interface Software}

To seamlessly simulate the post-synthesis netlists in the open source simulator Verilator, we develop an interface software that connects them to facilitate the tool automation. The interface software is a C++ program that performs four major tasks.

\begin{enumerate}

\item The first task is to open the exported netlist, add and/or remove some lines of Verilog code to obtain a synthesizable structural description that can be compiled by the Verilator tool. This task only includes the Verilog modules for the BELs in the structural description, generating the minimum necessary code to the Verilator compilation process. The generated Verilog file is going to be used in the fault injection campaign as the ``golden copy".

\item The second task is to generate the input stimuli that will be used by the Verilator testbench during the fault injection campaign. The input stimuli can be generated in two different ways. First, for an extensive fault injection campaign, it generates all combinations of input stimuli. Second, when the extensive campaign is not feasible due to the long simulation time, it generates a pseudo-random fault injection input stimuli. In the pseudo-random case, the C++ software uses the boost C++ library to generate the pseudo-random integer numbers in a uniform distribution that is converted to its binary correspondent.

\item The third task is to automatically generate the testbench that wraps the C++ model of the Verilog structural description. This testbench file, when executed, receives two parameters: the stimuli input file generated in the second task and the desired output file name to write the testbench results. The testbench algorithm can be summarized into three different stages. The first stage is the input stimuli read; the second is the input injection (including the clock event generation); and the last stage is reading the output from the Verilator simulator and writing the results to the output file. These three stages are repeated for the number of cycles intended for which the simulation runs.

\item The fourth task is to generate all the faulty copies of the golden circuit by inverting only the logic function programmed at each lookup-table (LUT) for each file. Each copy has only one faulty component and in this case, the total number of faulty copies is equal to the total number of LUTs. In our current tool, only one LUT function is inverted in each file. But it can be easily extended to support fault injection at multiple LUTs and other types of BELs, which is left for future work. 
\end{enumerate}

\subsection{Cycle-Accurate Fault Simulation}
Verilator is a cycle-accurate open source simulator \cite{OMS}. Compared to an event-driven simulator---such as Xilinx Vivado logic simulator, which executes processes based on event triggering---Verilator executes the whole design in a topological order at every simulation cycle. Thus, the number of simulator-specific variables is significantly reduced and the simulation time is much faster.
In our framework, Verilator takes the Verilog description and compiles the code into a much faster optimized and optionally thread-partitioned model. This model is then wrapped into a C++ module, all using parameters similar to GCC or Synopsys’s VCS that can be easily automated with Makefiles and the make command, available in any Unix and Unix-like operating systems.

\subsection{Vulnerability Score Histogram Generation}

To better guide the soft error vulnerability analysis, we also calculate a normalized vulnerability score for each LUT and generate a histogram of vulnerability scores for all the LUTs in the design. The vulnerability score for each LUT is calculated as follows. First, for each output bit of a LUT, we multiply 1) the error possibility based on simulation results by 2) its weight, which can be configured by users (the default weight value is 1 for all output bits). 
Second, we divide the sum of all products calculated in the first step by the sum of all weights of output bits to get the vulnerability score for a LUT. 
The histogram of vulnerability scores for all the LUTs demonstrate which LUTs are more vulnerable to the bit-flip errors. 

\section{Fault Simulation Tool Comparison}
\label{sec:comparison}

To demonstrate the performance gain of cycle-accurate simulation using Verilator in our framework, we compare it to the Xilinx Vivado ISIM simulator. Both simulators are evaluated with the widely used ITC'99 benchmark subset circuits \cite{ref:ITC99}. 
Both simulators are parallelized using the make command with the -j option and use 12 cores. 
Ideally, the each circuit should be simulated with all possible combinations of inputs. But as this number grows exponentially with the product of the number of input bits and the number of intended simulation cycles, we decided to use 1,024 pseudo-random input combinations due to the long time run. It already takes several hours to run the 1,024 input stimulus for the biggest circuit. Table \ref{tab:Performance} presents the simulation performance comparison for the Verilator and Vivado simulators. As shown in column 5, the cycle-accurate Verilator simulator is from 9.7 up to 53.8 times faster than the Vivado ISIM simulator. This is because the Verilator simulator compiles the Verilog code into a much faster optimized model and does not model intra-cycle events. 

\begin{table}[!ht]
\centering
\vspace{-0.05in}
\caption{Verilator and Vivado performance comparison}
\label{tab:Performance}
\begin{tabular}{lcccc}
\hline
        &             & \multicolumn{2}{c}{Time (s)} & Speedup of \\
Circuit & Num. of LUTs & Vivado   & Verilator &Verilator   \\
\hline
b01     & 5           & 17       & 0.45      & 37.8      \\
b02     & 4           & 16       & 0.56      & 28.6      \\
b03     & 23          & 50       & 1.49      & 33.6      \\
b04     & 85          & 200      & 7.59      & 26.4      \\
b05     & 109         & 276      & 13.75     & 20.1      \\
b06     & 8           & 22       & 0.73      & 30.1      \\
b07     & 63          & 152      & 5.13      & 29.6      \\
b08     & 18          & 46       & 3.51      & 13.1      \\
b09     & 34          & 72       & 1.47      & 49.0      \\
b10     & 34          & 85       & 1.58      & 53.8      \\
b11     & 82          & 194      & 7.13      & 27.2      \\
b12     & 213         & 609      & 37.7      & 16.2      \\
b13     & 45          & 102      & 3.16      & 32.3      \\
b14     & 1073        & 6219     & 642       & 9.7       \\
\hline
\end{tabular}
\vspace{-0.1in}
\end{table}

\section{Use Cases of Our Framework}\label{sec:results}

To demonstrate the usefulness of our framework, we conduct two case studies. 

\subsection{Case Study \#1: Cycle-Level Error Possibility}

In this case study, we demonstrate how one can use our framework to analyze the possibility that each LUT output bit can be affected by an input bit-flip error, as the time goes cycle by cycle. We choose this granularity because sometimes a certain output bit may represent vital output information, for example, whether the circuit is ready or not ('1' or '0').
To calculate the error possibility for each output bit of a LUT, we simulate the circuit with each faulty input combination. 
If there is an error in the output bit (comparing to the golden copy), its corresponding error counter (per output bit per cycle) is incremented by one. The corresponding error possibility is computed by dividing the total number of counted errors by the total number of simulation runs (i.e. input combinations). This process is repeated for every cycle. At the end, we generate an error possibility distribution for each LUT output bit cycle by cycle. 

As an example, let us consider the circuit b01 of the ITC'99 benchmark. 
Table \ref{tab:FItab} presents the detailed error possibility for each output bit of each LUT from cycle 1 to 4 by simulating the b01 circuit using our framework. 
Each row of the table presents the error possibility observed in each output bit of each LUT at each cycle. 
The 'Total' column presents the error possibility considering at least one error in the total run. 
The total error metric is adjustable so that application designers can create and utilize different approaches to calculate the total error value that are appropriate according to their  knowledge of the application.  

\begin{table}[ht]
\centering
\vspace{-0.05in}
\caption{Error possibility results for each LUT output bit from cycle 1 to 4 in the b01 circuit from ITC'99 benchmark}
\label{tab:FItab}
\begin{tabular}{lllllll}
\hline
 &         &\multicolumn{4}{c}{Cycle} &  \\
LUT & Output bit  & 1    & 2    & 3     & 4    & Total    \\
\hline
FSM\_st{[}0{]}& outp    &  0   & 0.25 & 0.38  & 0.44 & 0.25 \\
FSM\_st{[}0{]}& overflw & 0    & 0.25 & 0.25  & 0.22 & 0.25 \\ \hline
FSM\_st{[}1{]}& outp    & 0    & 0    & 0     & 0    & 0    \\
FSM\_st{[}1{]}& overflw & 0    & 0    & 0     & 0    & 0    \\ \hline
FSM\_st{[}2{]}& outp    & 0    & 1    & 0.5   & 0    & 1    \\
FSM\_st{[}2{]}& overflw & 0    & 0    & 0     & 0.62 & 0    \\ \hline
outp\_i\_1    & outp    & 1    & 1    & 1     & 1    & 1    \\
outp\_i\_1    & overflw & 0    & 0    & 0     & 0    & 0    \\ \hline
overflw\_i\_1 & outp    & 0    & 0    & 0     & 0    & 0    \\
overflw\_i\_1 & overflw & 1    & 1    & 1     & 1    & 1    \\
\hline
\end{tabular}
\end{table}

\begin{figure*}
  \vspace{0.05in}
  \includegraphics[width=\textwidth]{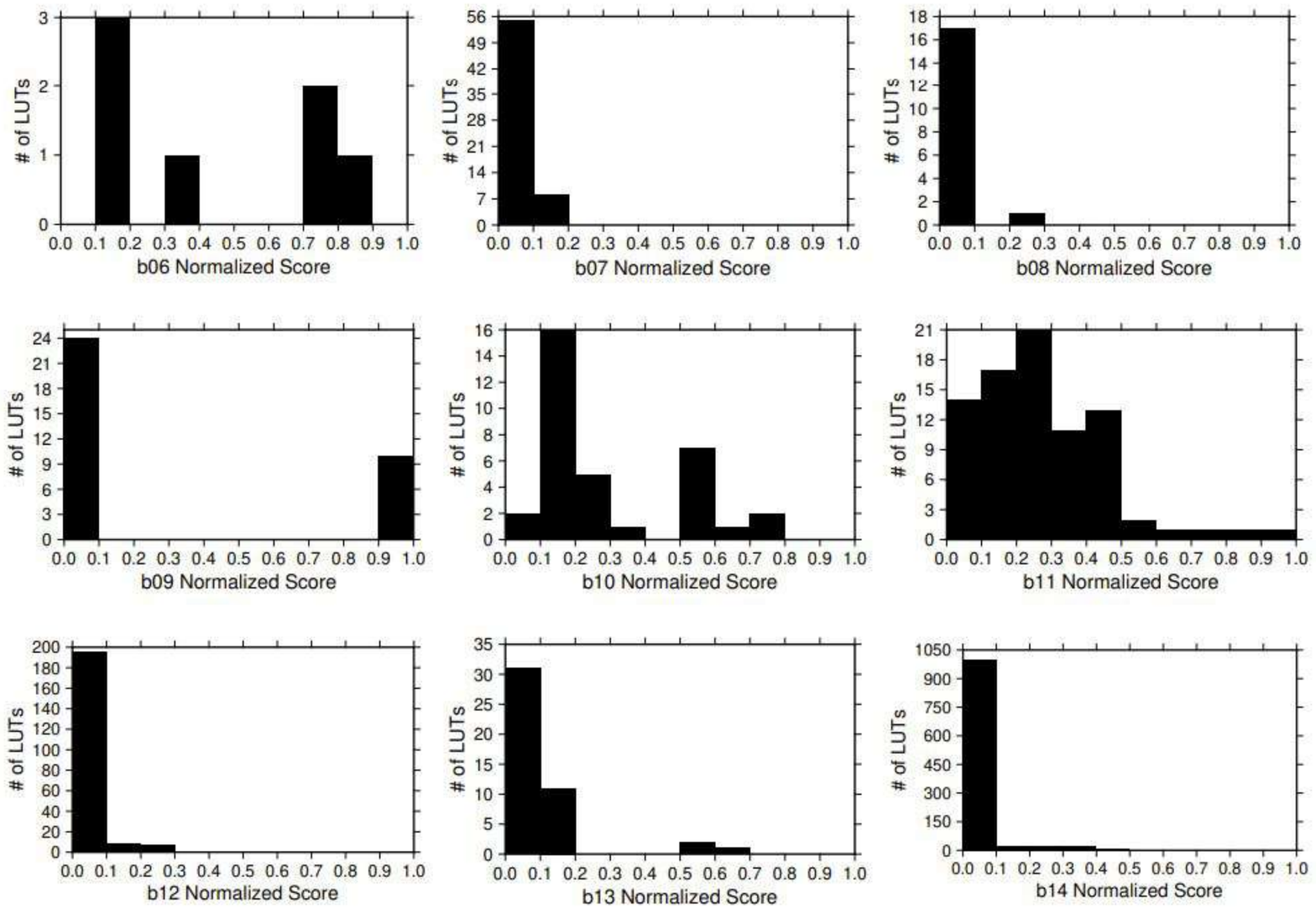}
  \caption{Histograms of the (normalized) vulnerability scores for the ITC'99 benchmark circuits.}
  \vspace{-0.1in}
    \label{fig:histograms}
\end{figure*}

As shown in Table~\ref{tab:FItab}, there is one LUT (FSM\_st{[}1{]}) that has no errors at the outputs. This is because the circuit topology has a reconvergence path. The fault propagates through two different paths and converges back to a specific component, where the bit flips twice resulting in an always masking condition. 
There are three LUTs that have a total error value of 100\%. Two of them (overflw\_i\_1 and outp\_i\_1) are very close to the outputs and the third is the most significant bit of the circuit FSM (Finite State Machine) state register (the FSM\_st{[}2{]}). 

Such analysis gives designers some guidance into which LUTs they should pay most attention to for better system reliability. This works well for small circuits that do not have too many LUTs, or a partial circuit that designers want to dive into for more details. Note that one can also perform such analysis per LUT, instead of per output bit of each LUT.



\subsection{Case Study \#2: Histogram of Vulnerability Scores}

In this case study, we demonstrate how one can use our framework to analyze which LUTs are more vulnerable to bit-flip errors in a circuit at a higher level, by analyzing the histogram of vulnerability scores of all LUTs. For illustrative purpose, we consider that all LUT output bits have the same importance (i.e., same weight 1) for the overall circuit functionality. We run the fault simulation for the ITC'99 benchmark subset circuits for 10 cycles to calculate the total error possibility for each LUT output bit as explained in case study \#1. 
Designers can configure the weights in our framework and decide how many cycles to simulate when they have better knowledge of the application. 


Figure~\ref{fig:histograms} presents the histograms for some circuits in the ITC'99 benchmark subset. Each sub-figure shows the number of LUTs (y-axis) that fit within a 0.1 normalized vulnerability score range (x-axis) for a circuit. 
As shown in Figure \ref{fig:histograms}, for some circuits, the majority of LUTs have normalized vulnerability scores below or equal to 0.5, e.g., circuits b07, b08, b11, b12, b13 and b14. Such circuits are less vulnerable to bit-flip
errors. For the small portion of LUTs that have normalized vulnerability scores above 0.5, one can afford to protect them using an expensive yet effective method like TMR (Triple Modular Redundancy) technique~\cite{1395771}. 
On the other hand, some circuits, such as b06, b09 and b10, have a significant amount of LUTs with normalized vulnerability score above 0.5. Such circuits are more vulnerable to bit-flip errors and need more protection.
In order to reduce the overhead of the error protection circuit, one may consider to apply the more lightweight DWC (Duplication With Comparison) technique \cite{1363710} or even redesign the circuit to reduce the number of vulnerable LUTs.

\section{Application for Large-scale Circuits}
The proposed fault simulation is not limited to small circuits, but it can be challenging to perform the fault simulation on large designs due to the computational complexity involved. Large FPGA designs can have thousands or even millions of BLEs, making it impractical to simulate every possible fault (even random faults) in the circuit.

To address this issue, partitioning techniques can be used to divide the large design into smaller, more manageable parts that can be simulated individually via the proposed accurate method. The design can be partitioned based on various criteria, such as functional blocks, critical parts, or input/output interfaces. Each partition can then be simulated separately using the proposed cycle-accurate fault simulation techniques, and the results can be combined to obtain the overall reliability of the design.
In addition, it can also help identify potential problems in specific parts (i.e. critical parts) of a large design, allowing for targeted optimization and testing.

\section{Conclusion}\label{sec:concl}

In this paper, we have presented a framework that enables fast and cycle-accurate analysis of soft error vulnerability for FPGA-based circuits in the early design stage. The fault model is at the basic element (BEL) level in an already optimized post-synthesis netlist. In our framework, a post synthesis netlist is first extracted for a given FPGA design or partial design. Then, bit-flip errors are inserted at BEL level. Both the golden copy and faulty copies run through the cycle-accurate simulator Verilator to identify the effects of faulty LUTs. 
By applying user-defined weights to each output bit, a histogram of vulnerability scores will be generated to guide designers to identify and rank the more vulnerable LUTs in the design.
Two case studies have been conducted on the widely used ITC'99 benchmark circuits to demonstrate the usefulness of our framework: one on cycle-level error possibility analysis for each LUT output, and the other on histogram of vulnerability scores of all LUTs. In general, our framework is up to 53x faster than the Xilinx Vivado simulation, while achieving cycle-level accuracy.


\bibliographystyle{IEEEtran}
\bibliography{biblo.bib}

\end{document}